# Optical Reading of Nanoscale Magnetic Bits in an Integrated Photonic Platform

Hamed Pezeshki, Pingzhi Li, Reinoud Lavrijsen, Jos J. G. M. van der Tol, and Bert Koopmans

*Abstract*— In this paper, we propose a compact integrated hybrid plasmonic-photonic device for optical reading of nanoscale magnetic bits with perpendicular magnetic anisotropy in a magnetic racetrack on top of a photonic waveguide on the indium phosphide membrane on silicon platform. The hybrid device is constructed by coupling a doublet of V-shaped gold plasmonic nanoantennas on top of the indium phosphide waveguide. By taking advantage of the localized surface plasmons, our hybrid device can enable detection of the magnetization state in magnetic bits beyond the diffraction limit of light and enhance the polar magneto-optical Kerr effect (PMOKE). We further illustrate how combining the hybrid device with a plasmonic polarization rotator provides magneto-optical read-out by transforming the PMOKE-induced polarization change into an intensity variation of the waveguide mode. According to the simulation results based on a three-dimensional finite-difference time-domain method, the hybrid device can detect the magnetization states in targeted bits in a magnetic racetrack medium down to ~ 100×100 nm², regardless of the magnetization state of the rest of the racetrack with a relative intensity contrast of greater than 0.5% for a ~ 200×100 nm² magnetic bit. We believe our hybrid device can be an enabling technology that can connect integrated photonics with nanoscale spintronics, paving the way toward ultrafast and energy efficient advanced on-chip applications.

*Index Terms*— Photonic integrated circuits, Plasmonics, Spintronics, Indium phosphide, Magneto-plasmonics, Polar magneto-optical Kerr effect

## I. Introduction

WITH the increase in demand for high bit-rate data transfer in the field of telecommunications and quantum information, the need for new technologies for high speed and reliable data reading and writing is foreseen. Advancements in the field of integrated photonics have enabled wide bandwidth optical data transmission on photonic integrated circuits, while spintronics has empowered high bit-rate data storage [1]. As a further upgrade, it is expected that a direct optical reading of spintronic domains, without any intermediate controlling high-frequency electronics, can bring higher bit-rate data transfer as well as enhance the energy efficiency of the communication platform [1], [2].

However, the drastic size mismatch between the diffraction-limited waveguide mode (> 400 nm [3]) and spintronic devices (< 150 nm [1]), makes photonic access of spintronics imcompatible. Moreover, the limited interaction cross-section results in low magneto-optical (MO) response.

Being able to concentrate the light energy without excessive photonic elements is the key to address the above mentioned problems. For over a couple of decades, it has been demonstrated that nanostructures made of noble metals like gold can confine the light energy in nanoscale spots at the metal-dielectric interface and magnify it as a result of the localized surface plasmon resonance [4], [5]. Due to the localized and enhanced electric field, the light-matter interaction within these nanostructures (plasmonic nanoantennas) gets boosted, which leads to an effectively larger cross-section of scattering and absorption powers of coupled nanoparticles. Nowadays, such unique features of plasmonic nanoantennas allow them to play a key role in addressing nanoscale elements in different areas such as telecommunication [6], [7] and biosensing [8], [9]. Hence, it has been anticipated that we can improve MO activity at the nanoscale based on magneto-plasmonic effects via incorporating plasmonic nanoantennas in our photonic device.

In this paper, we propose an integrated hybrid plasmonic-photonic device for enhanced optical reading of magnetic bits by taking advantage of magneto-plasmonic effects. The targeted functionality is to read out data, which are stored in the up and down magnetization of ferromagnetic bits of a racetrack with perpendicular magnetic anisotropy [10], placed on top of a photonic waveguide on the indium phosphide (InP) membrane on silicon (IMOS) platform. The read-out functionality is based on the polar magneto-optical Kerr effect (PMOKE), which manifests itself as the induced change of the polarization state of light reflected off the surface of a magnetic material depending on the magnetization state. The intrinsically small MO signal is enhanced by coupling a doublet of gold plasmonic nanoantennas with the racetrack on top of a photonic waveguide on the IMOS platform. This allows the incident light propagating through the waveguide to be focused efficiently on the targeted nanoscale magnetic bit in the coupled racetrack and to enhance the naturally weak PMOKE response by improving the MO interactions with the magnetic bits. To detect the magnetic states in a targeted magnetic bit, the resultant polarization change induced by PMOKE is transformed into an intensity variation of the transverse

H. Pezeshki, Pingzhi Li, Reinoud Lavrijsen, and Bert Koopmans are with the Department of Applied Physics, Eindhoven University of Technology, Eindhoven, 5612 AZ, Netherlands (email: h.pezeshki@tue.nl).

H. Pezeshki, Reinoud Lavrijsen, Jos J. G. M. van der Tol, and B. Koopmans are with the Eindhoven Hendrik Casimir Institute, Center for Photonic Integration, Eindhoven University of Technology, Eindhoven, 5600 MB, Netherlands.

magnetic ($TM_0$) waveguide mode [11] using our proposed plasmonic polarization rotator. Based on three dimensional finite-difference time-domain simulations, Lumerical, FDTD solver [12], we found that the presented hybrid device can enable the detection of the magnetization state in targeted bits in a magnetic racetrack medium down to a footprint of ~100×100 nm$^2$, without being disturbed by neighbouring bits. In contrast, for a photonic waveguide without the plasmonic nanoantennas, i.e. the bare waveguide device, this is not possible due to very weak PMOKE, while the extended waveguide mode would mix signals from neighbouring bits, ending up in ambiguous reading. Moreover, by using the polarization rotator, we show that the memory state in a magnetic bit with a size of ~200×100 nm$^2$ can be read-out optically with an intensity contrast of more than 0.5% only with using the hybrid device. The proposed device is a generic model which can be implemented in other photonic platforms such as silicon-on-insulator [13], [14] and silicon nitride [15], [16]. Based on our theoretical results, we believe that our proposed device concept can be an enabling technology which offers a method for direct optical reading of magnetic bits without intermediate electronics and it can be as well useful for magnet based sensing technologies.

## II. DESIGN STRUCTURE AND FUNCTION PRINCIPLE

The schematic diagram of the proposed integrated device for optical reading of magnetic bits in a racetrack is shown in Fig. 1(a). As depicted, the device consists of two sections: the reading module and the polarization rotator section. The whole device is based on the IMOS technology [17]. The reading module is composed of a doublet of V-shaped gold plasmonic nanoantennas coupled with a magnetic racetrack as a top-cladding on the InP waveguide. A magnetic racetrack enables densely storing of information as up and down magnetization states [18], which can be moved along the racetrack by electrical current [19]–[21]. The racetrack is modelled as a multilayer stack (from bottom to top) of 4 nm heavy metal tantalum seed layer, 2 nm platinium layer, 2 nm ferromagnetic cobalt layer with a MO Voigt constant of $Q = 0.154-0.100i$ [22] (responsible for PMOKE in our simulation model), and a 2 nm platinium capping layer. A continuous wave laser light source with a free space wavelength of $\lambda_0 = 1550$ nm is coupled to the transverse electric ($TE_0$) waveguide mode. Upon interaction between the $TE_0$ waveguide mode and the plasmonic nanoantenna under the resonance condition, the localized surface plasmons of the plasmonic nanoantenna get excited and enhance the electric field at its nanoscale hot spot, where the magnetic racetrack is coupled. The concentrated electric field leads to an enhanced PMOKE. As the pure $TE_0$ mode interacts with the magnetic cladding, partly a $TM_0$ mode with a small magnitude is created due to PMOKE, whose phase is magnetization dependent, i.e. changes by 180° when the magnetization reverses. As a result of the birefringence in the waveguide, the $TE_0$ mode and PMOKE-induced $TM_0$ mode beat along the propagation distance. Therefore, the rotation and phase of the this beating is also magnetization dependent. When the propagating light passes through the polarization rotator, this phase difference is transformed to an intensity variation of the emerged $TM_0$ mode due to the partial conversion of the $TE_0$ mode to $TM_0$ mode [11], by which the change in the magnetization state can be detected using a photodetector.

The design parameters are presented in Fig. 1. The width and height of the waveguide are $w_1 = 570$ nm and $h = 280$ nm. The plasmonic nanoantenna is based on a doublet of V-shaped gold nanoantennas, each of which is oriented at $\theta = 45°$ with reference to the waveguide direction and has a length, width and height of $l_{NA} = 120$ nm, $w_{NA} = h_{NA} = 30$ nm, where the subscript 'NA' is the acronym for nanoantenna. The geometry of the plasmonic nanoantenna is optimized to have a resonance peak at $\lambda_0 = 1550$ nm. The racetrack has a witdh and height of $w_{MR} = 120$ nm and $h_{MR} = 10$ nm, where the subscript 'MR' stands for magnetic racetrack. Figure 1(c) shows a schematic of the polarization rotator. A gold metal film with a thickness of 30 nm is stacked on top of the waveguide, where to minimize the absorption loss by the metal film, a silica ($SiO_2$) spacer layer with a thickness of 20 nm is placed in between the metal film and the waveguide. The length, width, and height of the polarization rotator are $l_{PR} = 1860$ nm, $w_{PR} = 180$ nm, and $h_{PR} = 50$ nm (where 'PR' stands for polarization rotator), respectively, in order to match the quarter of beating length and rotate the eigenmode by 45°. The width of the waveguide in the polarization rotator section is reduced to $w_2 = 440$ nm to maximize the $TE_0$ to $TM_0$ conversion efficiency. Note that the refractive indices of the materials used in the model are taken from the built-in library of the Lumerical, FDTD solver [12].

## III. RESULTS

In this section, we present the steps involved in the process of reading the magnetization state optically. The first step is based on PMOKE which is an induced change in the polarization state of the $TE_0$ mode due to the MO activity.

In the next step, we will convert the PMOKE-induced polarization change to an intensity variation of the $TM_0$ mode which can be easily detected using an on-chip photodetector.

### A. Optical Reading using PMOKE

By optically reading the magnetization using PMOKE, data stored in up and down magnetization states can be transferred to the photonic state. We assessed the performance of the optical reading using PMOKE based on two major factors. The first factor is the magnitude of the resultant PMOKE-induced polarization change (Kerr rotation), while the second one is the minimum footprint of the targeted magnetic bit in which the magnetization state can be identified regardless of all adjacent magnetic bits in the rest of the racetrack.

To investigate the minimum size of the targeted bit whose magnetization state can be determined, we perform a study of the evolution of the Kerr rotation as a function of the size of the targeted magnetic bit, along the light propagation direction inside the waveguide. This analysis is done in the presence of the oppositely magnetized rest of the racetrack (see the inset in Fig. 2(a)). Here, we study the Kerr rotation magnitude and phase as a function of the targeted bit size, by which the smallest readable bit size can be identified.

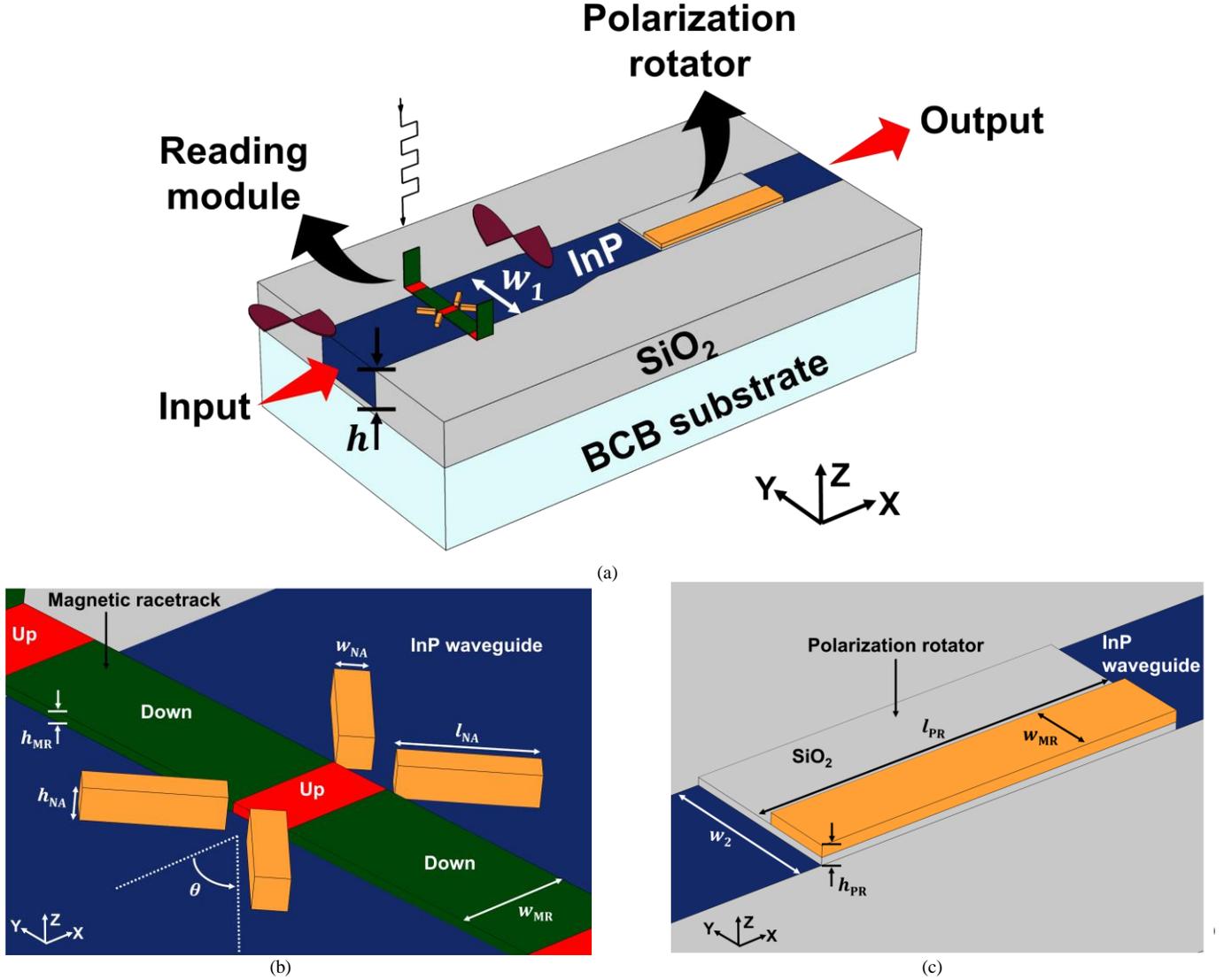

Fig. 1. The concept of the integrated device for all-optical reading of magnetic bits. (a) The schematic diagram of the device illustrating the operation principle. (b) The magnified view of the reading module showing the dimensions of the plasmonic nanoantenna elements and magnetic racetrack. The waveguide's width and height are $w_1 = 570$ nm and $h = 280$ nm (see (a)), the racetrack width and height are $w_{MR} = 120$ nm and $h_{MR} = 10$ nm, respectively; the length, width and height of the plasmonic nanoantenna elements are $l_{NA} = 120$ nm, $w_{NA} = h_{NA} = 30$ nm, and they are oriented at $\theta = 45°$ with reference to the waveguide direction. (c) The magnified view of the polarization rotator, where the width of the waveguide is $w_2 = 440$ nm, respectively; the length, width, and height of the polarization rotator are $l_{PR} = 1860$, $w_{PR} = 180$ nm, and $h_{PR} = 50$ nm, respectively. 'NA', 'MR', and 'PR' stand for nanoantenna, magnetic racetrack, and polarization rotator.

Figure 2(a) illustrates the evolution of the Kerr rotation magnitude as the $TE_0$ light mode propagates through the waveguide of the hybrid device, where the targeted bit has a domain width (DW) of 60 nm and is surrounded by the oppositely magnetized background (red domain in green background in the inset in Fig. 2(a)). As light interacts with the magnetic racetrack (orange region), we can clearly see a rise in the rotation of the polarization as a result of PMOKE. Here, we also note that the small polarization oscillation before the magnetic section is due to the light reflection from the plasmonic nanoantenna and the magnetic racetrack back to the input. Due to the birefringence in the waveguide, we have a beating between the $TE_0$ mode and the PMOKE-induced $TM_0$ mode which results in the oscillation of the Kerr rotation magnitude as light propagates through the waveguide.

By increasing the width of the targeted bit (see Fig. 2 (b)) and comparing the results, we can observe that not only the magnitude of the Kerr rotation changes but also the phase varies. This change is such that the Kerr rotation at DW = 60 nm is reversed by approximately 180° compared to DW widths of 200 and 570 nm. To explain the reason behind this reversal, it is important to note that the detected Kerr rotation is the superposition of the contributions of both the targeted red domain and oppositely magnetized green domains. When the targeted bit width is very small, e.g. DW = 60 nm, the sum of the Kerr rotations from the green domains dominate that of the targeted bit due to the small MO contribution from the targeted region. In this case, we cannot detect the magnetization state in the targeted bit unambiguously, since the outcome would depend on the content of neighboring bits. In contrast, when DW is either 200 nm or 570 nm (or any value in between), the Kerr rotation of the targeted bit becomes larger than the sum of the Kerr rotations of the oppositely magnetized regions. In this case, the magnetization state of the targeted magnetic bit can be uniquely identified.

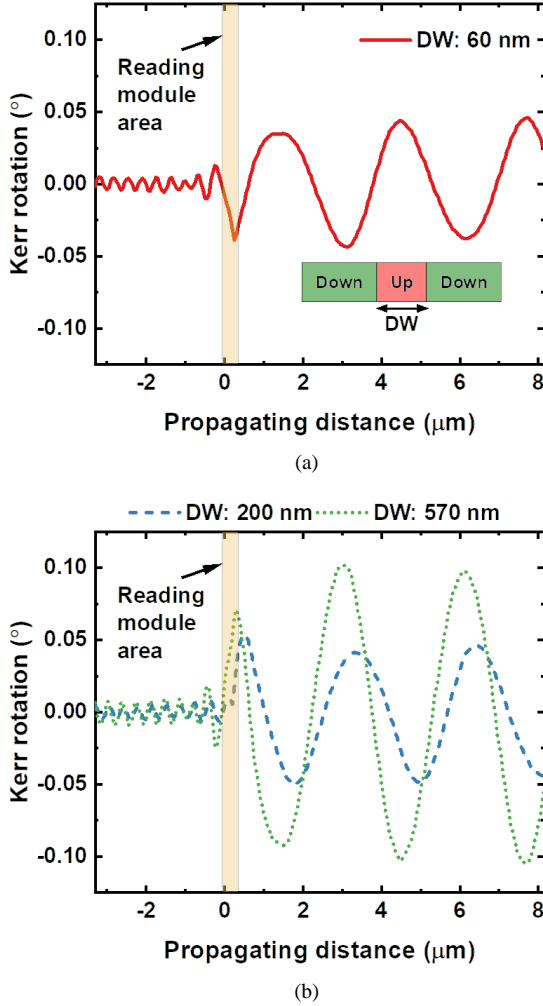

(a)

(b)

Fig. 2. The evolution of the PMOKE response for the hybrid device in terms of the different targeted domain width (DW) of: (a) 60 nm, and (b) 200 and 570 nm, respectively. The inset in (a) shows the targeted bit in red (Up magnetization) is surrounded by oppositely magnetized bits in green (Down magnetization).

Hence, we plotted the magnitude and phase of the Kerr rotation for both the bare waveguide and hybrid devices as a function of DW in Fig. 3. As shown, a sudden transition in the phase of the Kerr rotation is observed at DW for which the magnitude of the Kerr rotation has a minimum. The minimum Kerr rotation happens at DW = 120 nm (200 nm) for the hybrid (bare waveguide) device, which is accompanied by a jump in the phase of the Kerr rotation. For very small DWs, i.e. DW < 120 nm (200 nm) in the hybrid (bare waveguide) device, the PMOKE response of the targeted bit is weaker than the superposition of the rest of the bits in the racetrack due to the limited MO contribution from the targeted region. Thus, the oppositely magnetized bits in green determine the magnitude and phase of the Kerr rotation. On the other hand, for DW > 120 nm (200 nm), the PMOKE response from this bit (red domain) has become dominant. Thus, the magnetization state in the targeted magnetic bit can be explicitly identified above this value of DW, regardless of the magnetization state in the rest of the racetrack.

Based on the results, we can see that the minimum footprint for determining the magnetization state in the targeted bit is DW = 120 nm (200 nm) for the hybrid (bare waveguide) device, which is a measure for the resolution of the device. Comparison of the performance of the bare waveguide and hybrid devices indicates that the hybrid device enhances the resolution of magnetization read-out due to magneto-plasmonic effects beyond the diffraction limit. In general, this section illustrated the possibility of optically reading of the magnetization states in magnetic bits with subwavelength sizes of down to ~ 100 × 100 nm$^2$ (~ DW × $w_{MR}$), regardless of the magnetization state in the rest of the racetrack.

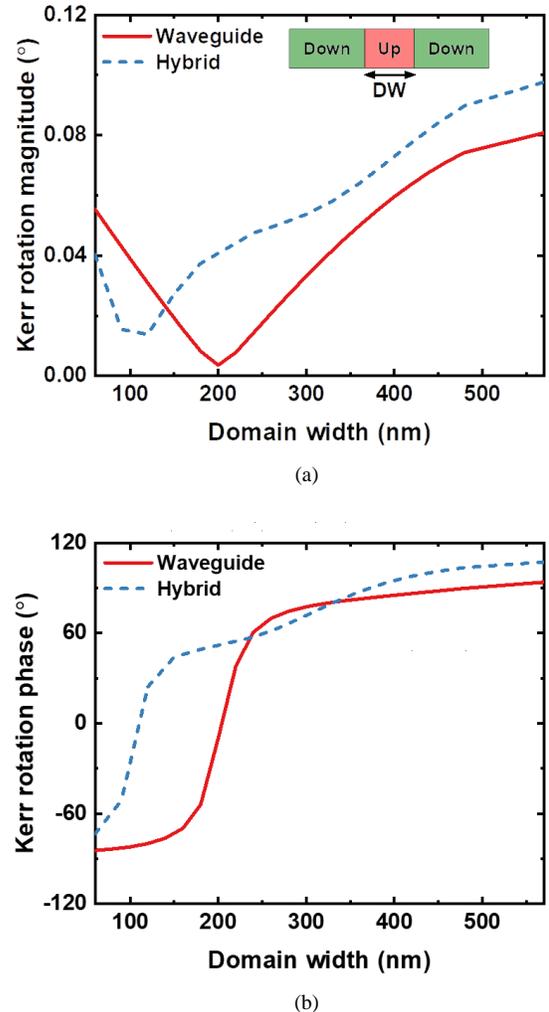

(a)

(b)

Fig. 3. The Kerr rotation magnitude and phase in terms of DW for the bare waveguide and hybrid devices, respectively.

## B. Detecting the Change in Magnetization State

As mentioned earlier, change in the magnetization state of the targeted bit induces a phase difference between the $TE_0$ and emergent $TM_0$ modes due to PMOKE. To be able to detect such a phase difference, we introduce a method for converting the phase difference to an intensity variation of the $TM_0$ mode [11] with the aid of a plasmonic polarization rotator [23]–[26] (see Fig. 1(c)). With using the polarization rotator, as light passes through it, due to the partial conversion of the $TE_0$ mode to the $TM_0$ mode, the phase difference between the $TE_0$ and $TM_0$ modes gets transformed into an intensity variation of the $TM_0$ mode. In this way, the magnetization state is encoded in the $TM_0$ intensity. As stated in section II, the polarization rotator section is comprised of a bilayer of $SiO_2$/gold (from bottom to top) which is asymmetrically positioned on top of the InP waveguide. The design parameters and schematic of the device are shown in Fig. 1(c).

To get insight about transforming a phase change to an intensity variation of the $TM_0$ mode using the polarization rotator, we use the polarization ellipse to illustrate the polarization states of the light mode with and without the polarization rotator. More explanation can be found in the previous work of our group [11]. The polarization state shown by the polarization ellipse can be quantified using the Stokes parameters $S_1$ to $S_3$ as follows [27]:

$$S_1 = \cos 2\varepsilon \cos 2\theta, \quad (1)$$
$$S_2 = \cos 2\varepsilon \sin 2\theta, \quad (2)$$
$$S_3 = \sin 2\varepsilon, \quad (3)$$

where $\theta$ and $\varepsilon$ are the polarization angle (Kerr rotation) and ellipticity angle (Kerr ellipticity), respectively. The Stokes parameter $S_1$ to $S_3$ show that whether the light mode is a pure $TE_0$ or $TM_0$ mode ($S_1$), an elliptically polarized mode ($S_2$) or a circularly polarized mode ($S_3$). Table I shows the values of Kerr rotation ($\theta$) and ellipticity ($\varepsilon$) at the output of the bare waveguide and hybrid devices in terms of the magnetization states at DW = 200 nm, with and without the polarization rotator. Based on Table I, the values of $\theta$ and $\varepsilon$ do not vary with the change in the magnetization state for the bare waveguide device with and without the polarization rotator. The reason is that at DW = 200 nm for the bare waveguide, the MO contribution is almost vanishing as shown in Fig. 3(a) which leads to $S_1 \approx 1$ and $S_{2,3} \approx 0$ for the cases of with and without the polarization rotator based on Eqs. (1) – (3). Figure 4(a) shows the polarization state of light (with some exaggeration for the sake of clarification) at the output of the hybrid device without the polarization rotator section for both up (red solid-line curve) and down (blue dashed-line curve) magnetization states, where two curves are overlapped. Based on the values of $\theta$ and $\varepsilon$ in Table I and Eqs. (1) - (3), the Stokes parameters are $S_1 \approx 1$ and $S_{2,3} \approx 0$ for both up and down magnetization states. In this case, the effect of PMOKE-induced polarization change is so small that it cannot alter the polarization state of the input mode significantly, and consequently we have a $TE_0$ light mode without the polarization rotator section. On the other hand, Fig. 4(b) shows the polarization ellipse for the hybrid device in the presence of the polarization rotator section which shows the two

TABLE I
KERR COMPONENTS, ROTATION ($\theta$) AND ELLIPTICITY ($\varepsilon$), FOR BARE WAVEGUIDE AND HYBRID DEVICES IN TERMS OF MAGNETIZATION STATES

| Kerr components | Bare waveguide | | | | Hybrid | | | |
|---|---|---|---|---|---|---|---|---|
| | Up (↑) | | Down (↓) | | Up (↑) | | Down (↓) | |
| | WoP[a] | WP[b] | WoP | WP | WoP | WP | WoP | WP |
| $\theta$ (°) | 0.003 | 53.4 | -0.003 | 53.4 | 0.04 | 53.3 | -0.04 | 53.5 |
| $\varepsilon$ (°) | 0.04 | 14.6 | 0.04 | 14.6 | 0.09 | 14.5 | 0.09 | 14.6 |

The values of $\theta$ and $\varepsilon$ are for the target domain witdh (DW) of 200 nm.
[a]WoP: Without the polarization rotator
[b]WP: With the polarization rotator

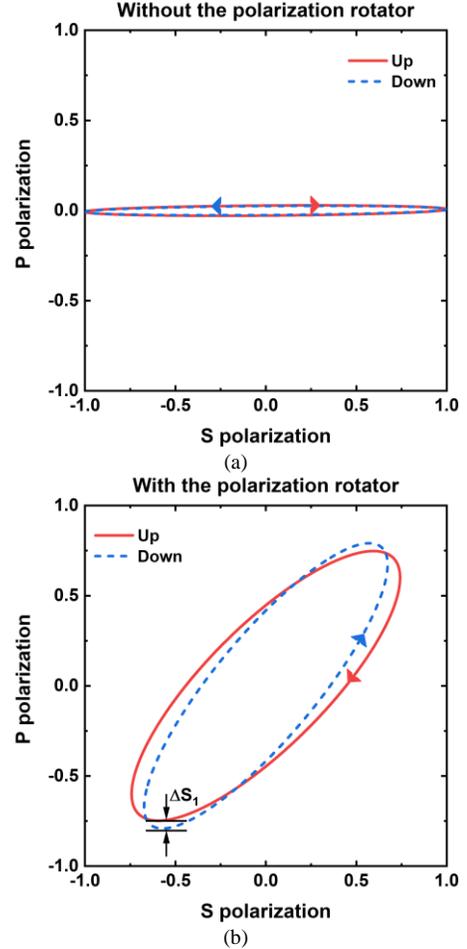

Fig. 4. Demonstration of transformation of the phase change to the intensity variation of a light mode propagating through the waveguide of the hybrid device using the polarization ellipse without and with the polarization rotator. The legends refer to up and down magnetization states. For the sake of clarity, the value of the magneto-optical Voigt constant, $Q$, is multiplied by 20.

curves do not completely overlap. According to Table I, the values of $\theta$ and $\varepsilon$ differ by 0.2° and 0.1°, respectively, when the magnetization state changes. Based on the values of $\theta$ and $\varepsilon$ in this case, $S_2 \approx 0.8$ and $S_3 \approx -0.5$ for both the magnetization states. In contrast, we have two slightly different values of $S_1^\uparrow \approx -0.251$ and $S_1^\downarrow \approx -0.256$ for up and down magnetization states, respectively. This difference in the values of $S_1$ originates from a variation in the intensity of the PMOKE-induced $TM_0$ mode, which explains the phase to intensity transformation that is used to read out the magnetic bits. Hence, we defined a figure of merit $\Delta S_1$, i.e. the relative contrast of the $TM_0$ mode, as follows:

$$\Delta S_1(\%) = \left|S_1^\uparrow - S_1^\downarrow\right| \times 100, \qquad (4)$$

Based on Eqs. (1) and (4), for the bare waveguide $\Delta S_1 \approx 0$. In contrast, our hybrid device offers $\Delta S_1$ of greater than 0.5% for the same targeted magnetic bit. One can see that using the bare waveguide, we cannot detect the change in the magnetization state in such a sub diffraction limit domain. However, our hybrid device with the help of magneto-plasmonic effects (offered by the proposed plasmonic nanoantenna) can overcome the diffraction limit and detect the magnetization change in a targeted bit size down to $\sim 100 \times 100$ nm$^2$ in the presence of oppositely magnetized neighboring bits.

## IV. Discussion

We numerically illustrated an ultracompact hybrid plasmonic-photonic device for optical read-out of the magnetization state in nanoscale magnetic bits. Using the hybrid device and a racetrack with a width of $\sim 100$ nm, the read-out of magnetic bits with a nanoscale size down to $\sim 100 \times 100$ nm$^2$ is possible using PMOKE, irrespective of the magnetization state in the rest of the racetrack. To complete the reading function, i.e. detecting the magnetization change on-chip, we proposed a method based on the polarization rotation principle. Our plasmonic polarization rotator transforms the PMOKE-induced polarization change in the TE$_0$ mode to an intensity variation of the TM$_0$ mode. Based on the simulation results, we showed that the hybrid device can detect the change in the magnetization state for a $\sim 200 \times 100$ nm$^2$ targeted magnetic bit with a relative contrast of greater than 0.5%, while the bare waveguide device is not able to detect.

During recent years, few groups have reported on enhancing PMOKE using plasmonic effects. Maccaferri *et al.* presented a periodic array of nickel nano-ellipsoids to enhance PMOKE using plasmonic resonance of this element [28]. It is noteworthy to mention that unlike gold, nickel cannot offer a strong plasmonic resonance as it suffers from a strong interband absorption at resonance condition [29]. Luong *et al.* demonstrated that in contrast to a titanium-cobalt composite nanohole array, a silver-cobalt composite nanohole array can provide larger PMOKE enhancement due to plasmonic effects offered by silver [30]. Freire *et al.* exhibited a PMOKE enhancement with growing a periodic 2D array of [cobalt/platinum]$_{10}$ (10 is the number of the repetition of the bilayer) on a gold metal layer [31]. In spite of reporting on improved PMOKE by these groups, the approaches introduced in these works are not ideally suitable for optically reading of nanoscale magnetic bits. In addition, from fabrication point of view, the integration of such structures to photonic integrated circuits is difficult. Very recently, our group demonstrated the first experimental report on the on-chip MO reading of a diffraction-limited magnetic bit with a cross section of $600 \times 400$ nm$^2$ and without the presence of oppositely magnetized nearby domains [11]. In contrast, our proposed device concept, by offering MO read-out beyond the diffraction limit, can play a key role in the realization of the future technology of hybrid spintronic-photonic memories with energy efficient switching and reading with high bit-rate data transfer and data storage capacity.

## V. Conclusion

We introduced an ultracompact integrated hybrid plasmonic-photonic device for optical reading of nanoscale ferromagnet bits having perpendicular magnetic anisotropy on the IMOS platform. The hybrid device, which is based on coupling a doublet of V-shaped gold plasmonic nanoantennas on top of the waveguide, strengthen the MO interaction beyond the diffraction limit of light with the help of magneto-plasmonic effects. According to the simulation results, the hybrid device can make possible the identification of the magnetization state for $\sim 200 \times 100$ nm$^2$ magnetic bits with a relative contrast of greater than 0.5%, but in general, targeted bits down to $\sim 100 \times 100$ nm$^2$ can be unambiguously detected irrespective of the magnetization state in the rest of the racetrack. We believe this device can have a potential impact on direct optical read-out and can encode information in the optical state.

## VI. Acknowledgment

This work is part of the Gravitation program 'Research Centre for Integrated Nanophotonics', which is financed by the Netherlands Organization for Scientific Research (NWO). This work is also part of the the European Union's Horizon 2020 research and innovation programme under Marie Skłodowska-Curie Grant Agreement No. 860060.

## References

[1] B. Dieny *et al.*, "Opportunities and challenges for spintronics in the microelectronics industry," *Nature Electronics*, vol. 3, no. 8. 2020. doi: 10.1038/s41928-020-0461-5.

[2] A. v. Kimel and M. Li, "Writing magnetic memory with ultrashort light pulses," *Nature Reviews Materials*, vol. 4, no. 3. 2019. doi: 10.1038/s41578-019-0086-3.

[3] F. J. Rodríguez-Fortuño, A. Espinosa-Soria, and A. Martínez, "Exploiting metamaterials, plasmonics and nanoantennas concepts in silicon photonics," *Journal of Optics (United Kingdom)*, vol. 18, no. 12. 2016. doi: 10.1088/2040-8978/18/12/123001.

[4] E. Hutter and J. H. Fendler, "Exploitation of localized surface plasmon resonance," *Advanced Materials*, vol. 16, no. 19, 2004, doi: 10.1002/adma.200400271.

[5] D. K. Gramotnev and S. I. Bozhevolnyi, "Plasmonics beyond the diffraction limit," *Nature Photonics*, vol. 4, no. 2. 2010. doi: 10.1038/nphoton.2009.282.

[6] W. O. F. Carvalho and J. Ricardo Mejía-Salazar, "Plasmonics for telecommunications applications," *Sensors (Switzerland)*, vol. 20, no. 9. 2020. doi: 10.3390/s20092488.

[7] H. Pezeshki, A. J. Wright, and E. C. Larkins, "Ultra-compact and ultra-broadband hybrid plasmonic-photonic vertical coupler with high coupling efficiency, directivity, and polarisation extinction ratio," *IET Optoelectronics*, 2022, doi: 10.1049/ote2.12063.

[8] J. R. Mejía-Salazar and O. N. Oliveira, "Plasmonic Biosensing Focus Review," *Chemical Reviews*, vol. 118, 2018.

[9] H. 'Pezeshki, "Lab-on-a-chip technology platform for biophotonic applications," PhD thesis, University of Nottingham, Nottingham, 2021.

[10] S. S. P. Parkin, M. Hayashi, and L. Thomas, "Magnetic domain-wall racetrack memory," *Science*, vol. 320, no. 5873. 2008. doi: 10.1126/science.1145799.

[11] F. E. Demirer *et al.*, "An integrated photonic device for on-chip magneto-optical memory reading," *Nanophotonics*, vol. 11, no. 14, pp. 3319–3329, Jun. 2022, doi: 10.1515/nanoph-2022-0165.

[12] "Lumerical Inc." https://www.lumerical.com/products/fdtd/, 2022.

[13] W. Bogaerts *et al.*, "Silicon microring resonators," *Laser and Photonics Reviews*, vol. 6, no. 1. 2012. doi: 10.1002/lpor.201100017.

[14] W. Bogaerts *et al.*, "Nanophotonic waveguides in silicon-on-insulator fabricated with CMOS technology," *Journal of Lightwave Technology*, vol. 23, no. 1, 2005, doi: 10.1109/JLT.2004.834471.

[15] A. Z. Subramanian *et al.*, "Low-Loss Singlemode PECVD silicon nitride photonic wire waveguides for 532-900 nm wavelength window fabricated within a CMOS pilot line," *IEEE Photonics Journal*, vol. 5, no. 6, 2013, doi: 10.1109/JPHOT.2013.2292698.

[16] K. Shang, S. Pathak, B. Guan, G. Liu, and S. J. B. Yoo, "Low-loss compact multilayer silicon nitride platform for 3D photonic integrated circuits," *Optics Express*, vol. 23, no. 16, 2015, doi: 10.1364/oe.23.021334.

[17] J. J. G. M. van der Tol *et al.*, "Indium Phosphide Integrated Photonics in Membranes," *IEEE Journal of Selected Topics in Quantum Electronics*, vol. 24, no. 1, 2018, doi: 10.1109/JSTQE.2017.2772786.

[18] R. Blasing *et al.*, "Magnetic Racetrack Memory: From Physics to the Cusp of Applications within a Decade," *Proceedings of the IEEE*, vol. 108, no. 8, 2020, doi: 10.1109/JPROC.2020.2975719.

[19] P. Li, T. J. Kools, R. Lavrijsen, and B. Koopmans, "Ultrafast racetrack based on compensated Co/Gd-based synthetic ferrimagnet with all-optical switching," Apr. 2022.

[20] K. S. Ryu, L. Thomas, S. H. Yang, and S. Parkin, "Chiral spin torque at magnetic domain walls," *Nature Nanotechnology*, vol. 8, no. 7, 2013, doi: 10.1038/nnano.2013.102.

[21] I. M. Miron *et al.*, "Perpendicular switching of a single ferromagnetic layer induced by in-plane current injection," *Nature*, vol. 476, no. 7359. 2011. doi: 10.1038/nature10309.

[22] Wouter Van Parys, "Optimization of an Integrated Optical Isolator Based on a Semiconductor Amplifier with a Ferromagnetic Metal Contact," 2009.

[23] L. Gao, Y. Huo, J. S. Harris, and Z. Zhou, "Ultra-compact and low-loss polarization rotator based on asymmetric hybrid plasmonic waveguide," *IEEE Photonics Technology Letters*, vol. 25, no. 21, 2013, doi: 10.1109/LPT.2013.2281425.

[24] L. Gao *et al.*, "On-chip plasmonic waveguide optical waveplate," *Scientific Reports*, vol. 5, 2015, doi: 10.1038/srep15794.

[25] J. Zhang, S. Zhu, H. Zhang, S. Chen, G. Q. Lo, and D. L. Kwong, "An ultracompact surface plasmon polariton-effect-based polarization rotator," *IEEE Photonics Technology Letters*, vol. 23, no. 21, 2011, doi: 10.1109/LPT.2011.2165206.

[26] D. Zhu, H. Ye, Z. Yu, J. Li, F. Yu, and Y. Liu, "Design of compact TE-polarized mode-order converter in silicon waveguide with high refractive index material," *IEEE Photonics Journal*, vol. 10, no. 6, 2018, doi: 10.1109/JPHOT.2018.2883209.

[27] E. Collett, *Field Guide to Polarization*. 2009. doi: 10.1117/3.626141.

[28] N. Maccaferri *et al.*, "Anisotropic Nanoantenna-Based Magnetoplasmonic Crystals for Highly Enhanced and Tunable Magneto-Optical Activity," *Nano Letters*, vol. 16, no. 4, 2016, doi: 10.1021/acs.nanolett.6b00084.

[29] F. Pineider and C. Sangregorio, "Nanomaterials for magnetoplasmonics," in *Novel Magnetic Nanostructures: Unique Properties and Applications*, 2018. doi: 10.1016/B978-0-12-813594-5.00006-0.

[30] H. M. Luong, M. T. Pham, B. Ai, T. D. Nguyen, and Y. Zhao, "Magnetoplasmonic properties of Ag-Co composite nanohole arrays," *Physical Review B*, vol. 99, no. 22, 2019, doi: 10.1103/PhysRevB.99.224413.

[31] F. Freire-Fernández, R. Mansell, and S. van Dijken, "Magnetoplasmonic properties of perpendicularly magnetized [Co/ Pt]N nanodots," *Physical Review B*, vol. 101, no. 5, 2020, doi: 10.1103/PhysRevB.101.054416.

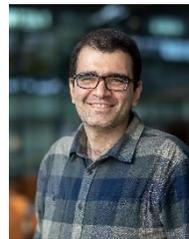

**Hamed Pezeshki** received his M.Sc from IAU, Science and Research Branch, Tehran, Iran in Feb. 2012. His research was in the field of photonic crystals with applications in telecommunications, for which he received a financial grant from the Iran Nanotechnology Initiative Council. Then, in Dec. 2016, he joined the Optics and Photonics group in the Department of Electrical and Electronic engineering at the University of Nottingham, Nottingham, United Kingdom, where he received the "Faculty of Engineering Research Excellence PhD Scholarship" for his course of study. His research was focused in the area of plasmonics with applications in biophotonics and telecommunications. Currently, he is a post-doctoral researcher in the Department of Applied Physics, Physics of Nanostructure group at the Eindhoven University of Technology. His current research is about the development of hybrid spintronic-photonic memories in collaboration with the Institute of Photonic Integration at the Eindhoven University of Technology. His research interests include photonic crystals, plasmonic nanostructures, and epsilon-near-zero materials.

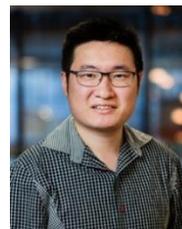

**Pingzhi Li** received his M.Sc from Catholic University of Leuven (KU Leuven) in 2019, where his research focus was on the characterization of the performance of integrated SOT-MRAM. He started his Ph.D (2019- ) at Eindhoven University of Technology as an early stage researcher funded by Marie Sklodowska-Curie Actions programme of the European Commision. During his Ph.D., his research focus is on the photonic control of spintronics.

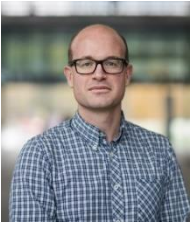

**Reinoud Lavrijsen** is an Assistant Professor in the Department of Applied Physics at Eindhoven University of Technology (TU/e). His areas of expertise include nanomagnetism and spintronics. Reinoud's research currently focuses on making the switching of nanomagnets energy efficient by combining different emerging physical principles in a single device. Fundamentally, this opens many exciting research opportunities, such as the complex interplay between the driving forces. Specifically, the group is currently studying unexplored territories of spintronics; control of magnetization-dynamics by using synthetic-multiferroic-heterostructures. Reinoud's objective is to obtain full control over the magnetization-dynamics of a (nano-) magnet without using magnetic fields. This is expected to result in a novel way of power-efficient and fast-coherent-control of magnetization. The potential for creating commercial devices (memory, data-storage, logic) is huge and may create new paradigms in the fundamentals of the underlying physics by new ways of probing competing interactions in a cleverly chosen, simple materials/device system.

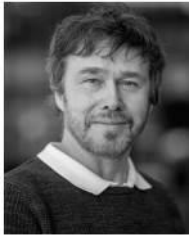

**Jos J. G. M. van der Tol** received the M.Sc. and Ph.D. degrees in physics from the State University of Leiden, Leiden, The Netherlands, in 1979 and 1985, respectively. In 1985, he joined KPN Research, where he had been involved in the research on integrated optical components for use in telecommunication networks. Since July 1999, he has been an Associate Professor with the Eindhoven University of Technology, Eindhoven, The Netherlands, where his research interests include opto-electronic integration, polarization issues, photonic membranes, and photonic crystals. He has coauthored more than 250 publications in the fields of integrated optics and optical networks and has 25 patent applications to his name. He has been working on guided wave components on III–V semiconductor materials. He has also been active in the field of optical networks, focusing on survivability, introduction scenarios, and management issues. His research interests include modeling of waveguides, design of electro-optical devices on lithium niobate, and their fabrication.

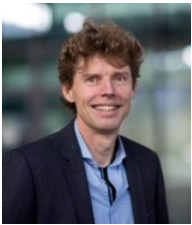

**Bert Koopmans** (1963) graduated and received his PhD degree at the University of Groningen. After a short stay as a postdoc at the Radboud University Nijmegen, he spent three years as a Humboldt Fellow at the Max-Planck Institute for Solid State Physics in Stuttgart. In 1997 he joined the Eindhoven University of Technology, where since 2003 he chairs the Group Physics of Nanostructures. His current research interests encompass spintronics, nanomagnetism and ultrafast magnetization dynamics. He participates in the NWO Gravitation program on integrated nanophotonics, initiating research on hybrid spintronic-photonic devices. In 2020 he was elected Distinguished Lecturer of the IEEE Magnetics Society.